\input{aipcheck}

\def\be{\begin{equation}}
\def\ee{\end{equation}}
\def\bea{\begin{eqnarray}}
\def\eea{\end{eqnarray}}

\def\gev{\, {\rm GeV}}

\newcommand{\gsim}{\lower.7ex\hbox{$\;\stackrel{\textstyle>}{\sim}\;$}}
\newcommand{\lsim}{\lower.7ex\hbox{$\;\stackrel{\textstyle<}{\sim}\;$}}

\newcommand{\pb}{\rm pb}

\newcommand{\cm}{\rm cm}

\documentclass[
    ,final            % use final for the camera ready runs
  ]
  {aipproc}

\usepackage{graphicx}
\usepackage{enumerate}

\layoutstyle{8x11single}

\begin{document}

\title{Asymmetric Dark Matter}

%95.35.+d Dark matter

\author{Jason Kumar}{
  address={Department of Physics and Astronomy, University of
Hawai'i, Honolulu, HI 96822, USA}
}

\begin{abstract}
We review the theoretical framework underlying models of
asymmetric dark matter, describe astrophysical constraints which
arise from observations of neutron stars, and discuss the prospects
for detecting asymmetric dark matter.
\end{abstract}

\maketitle

%%%%%%%%%%%%%%%%%%%%%%%%%%%%%%%%%%%%%%%%%%%%
%% MAINMATTER
%%%%%%%%%%%%%%%%%%%%%%%%%%%%%%%%%%%%%%%%%%%%

\section{Introduction}

Dark matter accounts for roughly 80\% of the matter in the universe.
Dark matter models which attempt to naturally explain this dark matter
density generally rely on one of two ``coincidences."    WIMP models
(or WIMPless variants~\cite{WIMPless}) rely on the fact that, for certain motivated choices
of dark matter particle mass and coupling, the relic density~\cite{Zeldovich:1965} which one would
calculate from thermal freeze-out is approximately that required by observation.
This coincidence is sometimes referred to as the ``WIMP Miracle."
Asymmetric dark matter relies on the second coincidence, namely, that the
dark matter and baryonic matter densities are similar (they differ by a
factor of $\sim 4$).

The basic idea of asymmetric dark matter is that the dark matter particle is
distinct from the anti-particle, and the current abundance arises almost
entirely from one species (conventionally taken as the particle).  In this way,
dark matter is similar to baryonic matter.  A variety of mechanisms have been
suggested for generating the asymmetry between baryonic matter and anti-matter.
If the same mechanism also generates the asymmetry between dark matter and
dark anti-matter, then the number density of dark matter should be simply related
to the baryonic number density.  If the mass of the dark matter particle is also similar to
mass of the nucleon ($\sim 1~\gev$), then the dark matter and baryon energy densities
will be similar.  For recent reviews of asymmetric dark matter, see~\cite{Petraki:2013wwa,Zurek:2013wia}.

The two generic features of asymmetric dark matter models are
\begin{itemize}
\item{dark matter annihilation is suppressed because only the particle is
abundant in nature, while the anti-particle is not,}
\item{the dark matter particle is light, with a mass similar to that
of the nucleon (though there are exceptions~\cite{Buckley:2010ui} to this result).}
\end{itemize}
In particular, many asymmetric dark matter models may thus
be able address recent hints for low-mass dark matter arising from the DAMA~\cite{Bernabei:2010mq},
CoGeNT~\cite{Aalseth:2010vx,Aalseth:2012if}, CRESST~\cite{Angloher:2011uu} and
CDMS~\cite{Agnese:2013rvf} experiments.

In these proceedings, we will discuss the theoretical motivations for these features,
the classes of models which satisfy the needed criteria, and the impact of these features on
asymmetric dark matter constraints and detection possibilities.

\section{Theoretical Framework}

\paragraph{Symmetries} An asymmetric dark matter candidate must be a
particle excitation of a complex field, for example, a complex scalar or a
Dirac fermion.  Only in this case is the particle distinct from the
anti-particle.  Generically, one expects that this should only be the case
if the field is charged under an unbroken $U(1)$ symmetry (or a subgroup of
$U(1)$ which rotates the field by a nontrivial complex phase).  The charge of the field
under this ``complexifying" symmetry is what distinguishes the particle from
the anti-particle.  Another way of seeing this is to note that a complex field
will have two degenerate mass eigenstates which correspond the particle and the
anti-particle.  The complexifying symmetry forbids any terms in the
Lagrangian (such as a Majorana mass term) which could split these eigenstates.
In the absence of this symmetry, one would generically expect the presence
of terms in the Lagrangian  which break the degeneracy, resulting in two non-degenerate
mass eigenstates, each of which is its own anti-particle.

If dark matter is truly stable, then generically it should be the lightest particle
charged under some unbroken symmetry.
Although this ``stabilizing" symmetry may be the same as the complexifying
symmetry described above, it does not have to be.  Importantly, the complexifying symmetry
cannot be a $Z_2$ parity, and dark matter need not be the lightest particle
charged under it.  Dark matter must be the lightest particle charged under
the stabilizing symmetry, and that symmetry may be a parity.

\paragraph{Self-annihilation}
The expectation for asymmetric dark matter is that there is no self-annihilation.
Particles can thus only annihilate against anti-particles, which are not abundant
in nature.  It is easy to see the origin of this expectation; a particle/anti-particle
system has no conserved charge which could obstruct the annihilation process,
while a particle/particle system has a conserved charge provided the particle is charged
under a symmetry which is not a parity.
Since asymmetric dark matter is charged under the complexifying symmetry, which is necessarily
not a parity, two asymmetric dark matter
particles cannot annihilate except to a lighter state with the same charge.

But there is a caveat to this argument.  Since dark matter need
not be the lightest particle charged under the complexifying symmetry, there is no
a priori reason why there should not be a lighter charged state to which two asymmetric
dark matter particles could annihilate.
Asymmetric dark matter self-annihilation is not forbidden unless the stabilizing symmetry
is the same as the complexifying symmetry.  If the stabilizing symmetry is instead a parity,
then charge under the stabilizing symmetry is only conserved modulo 2; although
dark matter decay is forbidden, annihilation would be allowed.

A simple example of this issue, familiar from the Standard Model, is the proton.
If we ignore the internal structure of the proton and treat it as a fundamental
particle, then the proton is distinct from the anti-proton because of its charge
($+1$) under $U(1)_{EM}$.  But this complexifying symmetry is not the stabilizing
symmetry, since the proton is not the lightest particle charged under $U(1)_{EM}$
(this is $e^\pm$).  The proton is the lightest particle with baryon number, and
its annihilation is forbidden if this stabilizing symmetry is also a continuous
complexifying symmetry, $U(1)_B$.
But if baryon and lepton number were just parity symmetries, then the proton would still be
stable, but the process $pp \rightarrow e^+ e^+$ would be allowed.

If asymmetric dark matter can self-annihilate with a large enough cross section,
then the asymmetry can be washed out.  In order for asymmetric dark matter to remain
asymmetric, its self-annihilation cross section must be small enough to have frozen
out in the early universe, resulting in an asymmetry which persists to the current epoch.
Such a small self-annihilation cross section could arise if the stabilizing symmetry
were a continuous symmetry which was very weakly broken to a parity.  An alternative approach
would be for the complexifying symmetry to be very weakly broken, thus implying that
the dark matter was ``almost" complex~\cite{Tulin:2012re}.

\paragraph{Particle Mass}  Although asymmetric dark matter does not in principle require a connection
between the dark matter and baryon asymmetries, much of the motivation is lost in the absence of
this connection.  We will thus focus on models for which these asymmetries are related.
For such models, it is worth noting that asymmetric dark matter really requires
two coincidences.  In addition to a mechanism connecting the net dark matter number density to the
net baryon number density, one requires a mechanism for relating the dark matter particle mass to
the mass of the lightest baryon ($\sim 1~\gev$).

\section{Classes of Models}

For a model of asymmetric dark matter, it is not enough for the particle and
anti-particle to be distinguishable; one must also have mechanism for generating
the dark matter asymmetry.  As with the baryon asymmetry, this amounts to satisfying
the Sakharov conditions.
There are a vast array of asymmetric dark matter models~\cite{Buckley:2010ui,ADM_models,MirrorMatter,Dutta:2010va} which
satisfy these conditions in a variety of ways.

One loose way of classifying asymmetric dark matter models is by the method for relating
the dark matter asymmetry to the baryon asymmetry.
Either the dark matter asymmetry could be generated first
and then transferred to baryons in some way, or vice versa, or both asymmetries could be
generated at the same time by the same mechanism.
These classifications are of course somewhat ambiguous, since the definition of
dark sector vs.~visible sector is sometimes just a matter of taste.
One can write a general template for
creating a model of asymmetric dark matter as follows:
\begin{itemize}
\item{{\it Pick a class:} decide if an asymmetry is first generated in the dark sector,
in the Standard Model sector, or in both simultaneously;}
\item{{\it Pick a generation mechanism:} choose a mechanism for satisfying the
Sakharov conditions and generating the initial asymmetry in the chosen sector;}
\item{{\it Pick a transfer mechanism:} choose a mechanism for transferring the
asymmetry, if needed, from the sector where it was generated to the other sector; }
\item{{\it Pick a mass:} find a mechanism for setting the dark matter particle
mass so as to generate the correct dark matter density.}
\end{itemize}
For each of these steps, a variety of possibilities have been studied.
Regardless of which sector contains the initial asymmetry, the generation of
this asymmetry requires that the Sakharov conditions be satisfied.  Mechanisms
for generating the initial asymmetry are thus typically generalizations
of standard mechanisms for baryogenesis/leptogenesis, and include
strongly first-order phase transitions, out-of-equilibrium annihilation or decay, the Affleck-Dine
mechanism~\cite{Affleck:1984fy}, etc.

The asymmetry can then be transferred from one sector to
another by sphalerons, annihilations, co-annihilations, decays, etc.  If the asymmetry
is being transferred to the Standard Model sector, then this mechanism must violate
$B$ or $L$.  If it is transferred to the dark sector, then it must violate $D$,
where $D$ is a dark sector quantum number which counts the asymmetry between dark particles
and dark anti-particles.  But this transfer mechanism need
not violate $CP$ or include an intrinsic departure from thermal equilibrium, since these conditions
have already been achieved with the generation of the initial asymmetry.

Finally, there are some models for which the dark matter mass can naturally be tied to the
GeV scale, thus explaining the relic density.  For example, for models of mirror matter~\cite{MirrorMatter},
the dark sector is a mirror copy of the Standard Model sector.
In this case, the dark matter candidate is a mirror baryon, and its mass is automatically
of the same scale as the Standard Model baryons.
In models where asymmetric dark matter is generated from
Hidden Sector Baryogenesis~\cite{Dutta:2006pt,Dutta:2010va}, the dark matter
candidate is chirally charged under a symmetry $U(1)_{T3R}$, under which the right-handed $b$- and $c$-quarks
and $\tau$-lepton are also charged.  Since all of these fields are chiral under $U(1)_{T3R}$,
all of their masses are proportional to the symmetry-breaking scale of $U(1)_{T3R}$.  The mass
of $b$, $c$ and $\tau$ are all ${\cal O}(1-10)~\gev$, implying that the symmetry-breaking scale of
$U(1)_{T3R}$, and thus the mass of the dark matter, are similar.

\section{Constraints and Detection Possibilities}

From the point of view of detection, the main constraints on asymmetric dark matter
arise from its relatively low mass and small self-annihilation cross section.  We can summarize
the difference between asymmetric dark matter and WIMPs, in this context, as follows:
\begin{itemize}
\item{{\it Direct Detection:} similar to low-mass WIMP searches.  Current
sensitivity is greatly degraded for the lighter asymmetric dark matter candidates;}
\item{{\it Indirect Detection:} no signals expected.  A self-annihilation cross section
large enough to be detected would erase the dark matter asymmetry;}
\item{{\it Collider Searches:} similar to low-mass WIMP searches.  Sensitivity
improves as the dark matter mass decreases;}
\item{{\it Astrophysical constraints:} asymmetric dark matter which is captured in large astrophysical objects
can collapse to form a black hole.  Constraints can be placed on the dark matter capture
rate by the observation of old neutron stars which have not collapsed.}
\end{itemize}

\subsection{Constraints from Old Neutron Stars}

An interesting new feature arising for asymmetric dark matter is the possibility of
constraining dark matter-nucleon interactions based on the absence of black hole formation
within large astrophysical bodies.  The basic idea is a variant of the dark matter search strategy used
by neutrino detectors.  When a dark matter particle scatters off a nucleus in any dense astrophysical body
(such as the sun), it will lose kinetic energy to the nucleus elastic recoil.  If the dark matter
particle velocity falls below the escape velocity of the astrophysical body, it will be gravitationally
captured.  After many orbits the dark matter particle will have scattered enough to thermalize, and the
dark matter will collect in a dense region near the core of the astrophysical body.

Neutrino detectors search for the neutrinos which are emitted from the sun or earth
when dark matter in this dense region annihilates.  But what if dark matter annihilation
were highly suppressed, as in the case of asymmetric dark matter?  In this case, dark matter would
keep collecting within the astrophysical body, with no mechanism for depleting the dark matter.
If enough dark matter is captured, the dark matter would become gravitationally unstable to collapse
into a black hole.  If this black hole grows, it could then destroy the astrophysical body.  Observations
of astrophysical bodies which have not been destroyed thus place a bound on the dark matter capture rate,
and in turn on the dark matter-nucleon scattering cross section.

Several authors have considered this class of constraints~\cite{NSbounds,Bramante:2013hn,Bell:2013xk} for neutron stars in globular clusters.  The
tightest constraints arise for bosonic asymmetric dark matter, for which there is no Fermi degeneracy
pressure to obstruct dark matter collapse.  We can briefly summarize the path of this type of analysis:
\begin{enumerate}[(1)]
\item{{\it Dark matter accumulates:} the accumulation rate depends on $\sigma_{nX}$ and $\langle \sigma_A v \rangle$,
the dark matter-neutron
scattering cross section and the thermally-averaged dark matter self-annihilation cross section, respectively.}
\item{{\it Dark matter thermalizes:} if the lifetime of the neutron star is not sufficient for
thermalization, then there is no constraint.}
\item{{\it Dark matter forms a Bose-Einstein Condensate:} for the relevant parameter space, when sufficient dark matter collects
at the neutron star core, it will form a Bose-Einstein condensate (BEC).}
\item{{\it Dark matter in the BEC phase crosses the bosonic Chandrasekhar bound:}  this bound depends on the dark matter
mass, $m_X$, and the strength of self-interactions.  Once dark matter crosses this bound, a black hole will
form. }
\item{{\it The black hole evolves:} a large black hole will grow by accreting baryonic and dark matter.  A small
black hole will quickly evaporate away.  If the black hole will grow large enough to consume the astrophysical
body within its lifetime, the point in parameter-space would be ruled out by observation.}
\end{enumerate}

We consider an asymmetric dark matter field $\phi$ with self-annihilation cross section $\sigma_A$ and a
repulsive $\lambda |\phi^2|^2/4!$ self-interaction.  This interaction is not generically forbidden by any symmetry of the theory,
so one does not expect $\lambda$ to be particularly small~\cite{Bramante:2013hn,Bell:2013xk}.\footnote{Note, such
a quartic interaction can be forbidden in some supersymmetric theories, but will in general be generated by terms
which break supersymmetry~\cite{Bell:2013xk}. }
We will take the dark matter density within a globular
cluster, $\rho_X$, to be $10^3~\gev /\cm^3$, as a benchmark.  The actual density may be much smaller, but this
would just result in a rescaling of the constraint on $\sigma_{nX}$.

We can express the Chandrasekhar bound for self-interacting bosonic matter as~\cite{BosonicChandrasekharBound}
\bea
N_{Chand} &\approx& {2m_{pl}^2 \over \pi m_X^2} \left(1 + {\lambda \over 32\pi} {m_{pl}^2 \over m_X^2} \right)^{1\over 2} .
\eea
A black hole will form when the number of particles in the BEC phase exceeds $N_{Chand}$.
If $\lambda$ is extremely small, then $N_{Chand} \propto (m_{pl} / m_X)^2$, but for natural values of $\lambda$
we instead find $N_{Chand} \propto (m_{pl} / m_X)^3$, similar to the fermionic Chandrasekhar bound.

$N_{acc}$, the number of dark matter particles accumulated within the core of the neutron star, is given by
\bea
N_{acc} &\sim& \sqrt{C_X V_{th} \over \langle \sigma_A v \rangle } \tanh \left[\sqrt{C_X \langle \sigma_A v \rangle \over V_{th} }
t_{ns} \right]
\eea
where $V_{th}$ is the volume of the thermalization region, $t_{ns} \sim 10~{\rm Gyr}$ is the neutron star lifetime and $C_X$ is the
capture rate.  The capture rate is proportional to $\rho_X \sigma_{nX}$~\cite{CaptureRate} provided $\sigma_{nX} \leq \sigma_{sat.}\sim
2.1 \times 10^{-9}~\pb$.  A bound on $N_{acc}$ can thus be rephrased as a bound on $\sigma_{nX}$, which can be compared to
results from direct detection experiments.  But it is important to note that, for $\sigma_{nX} \geq \sigma_{sat}.$, all dark
matter particles which reach the neutron star scatter against it; as a result, increasing $\sigma_{nX}$ cannot increase the
capture rate any further.  Thus, if dark matter is not excluded for $\sigma_{nX} \leq \sigma_{sat.}$, then there exists no
value of $\sigma_{nX}$ for which it is excluded.

\begin{figure}[!ht]
\centering
\includegraphics[scale=.75]{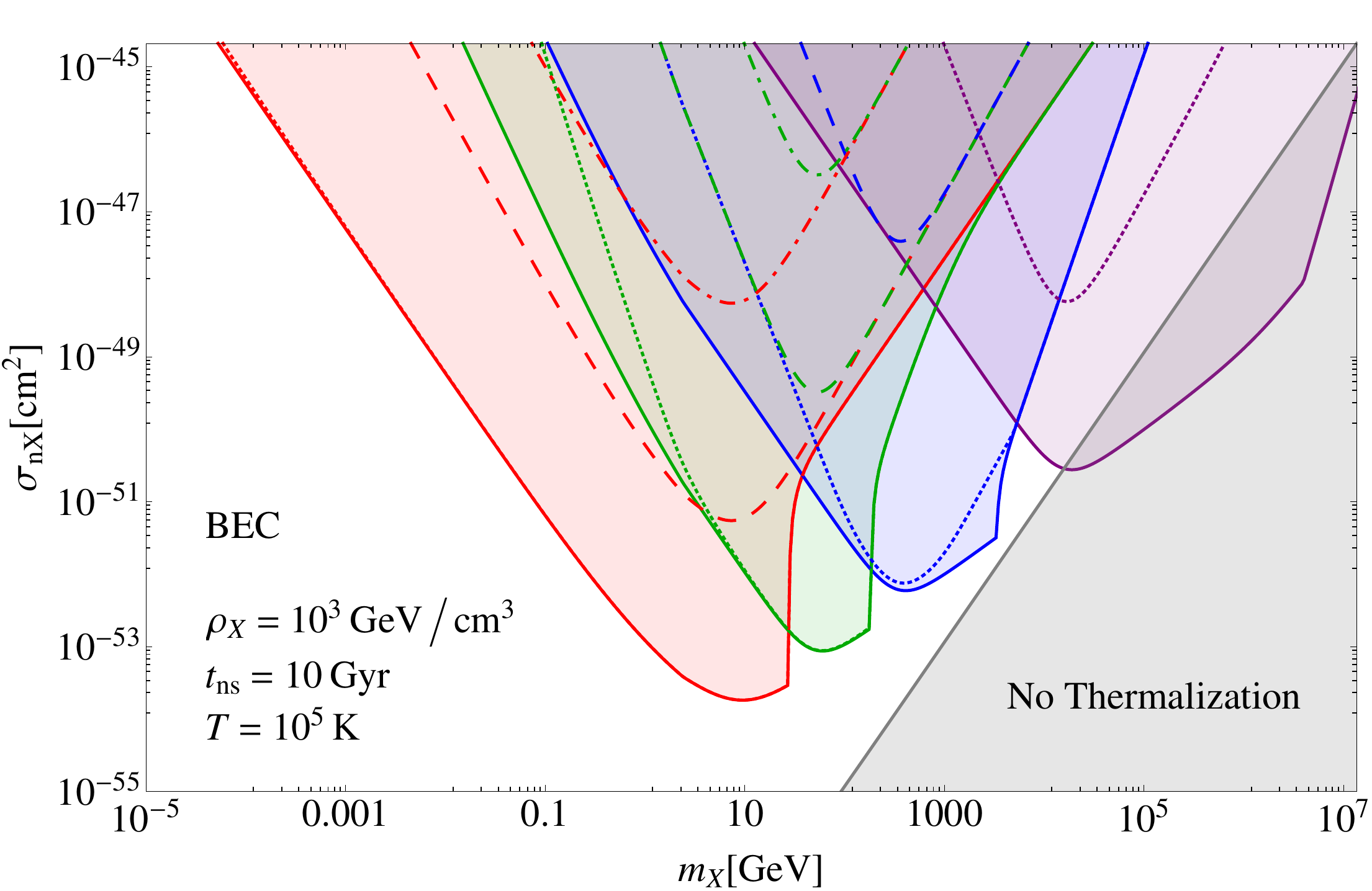}
\caption{Constraints on bosonic asymmetric dark matter, in the $(m_X, \sigma_{nX})$ plane, arising from observations of old neutron stars in globular clusters.
We assume a globular cluster density $\rho_X \sim 10^3 ~ {\rm GeV/cm^3}$.
The red, green, blue, and purple contours (from left to right) denote $\lambda = \{0,10^{-30},10^{-25},10^{-15}\}$, respectively.
Solid, dotted, dashed, and dot-dashed contours denote
self-annihilation cross sections $\langle \sigma_a v \rangle = \{0,10^{-50},10^{-45},10^{-42}\}~{\rm cm^3 / s}$, respectively.  In the gray region,
dark matter does not thermalize within the lifetime of the neutron star.
}
\label{fig:NeutronStarBounds}
\end{figure}

In figure~\ref{fig:NeutronStarBounds}~\cite{Bramante:2013hn}, we plot bounds on $\sigma_{nX}$ arising
from observations of old neutron stars in globular clusters
for a variety of choices for $\langle \sigma_A v \rangle$ and $\lambda$.  One can see that as $m_X$ increases, constraints
on $\sigma_{nX}$ initially become tighter because fewer dark matter particles are needed for black hole formation.  But for large
enough $m_X$, constraints on $\sigma_{nX}$ become very weak because the black hole which is formed is so small that it
evaporates away unless the black hole can capture dark matter rapidly enough to ``feed" the black hole and keep it growing.
For this reason, an extremely small self-interaction coupling $\lambda$ can cause these constraints to be slightly tighter;
self-interaction causes a the formation of a larger black hole, which can then grow to consume the neutron star.  But for even a
moderate value of $\lambda$, a black hole will never form.  Similarly, for $\langle \sigma_A v \rangle \gsim 10^{-12}~\pb$,
dark matter depletion will be rapid enough that a black hole will never form.  Thus, even a self-annihilation cross section
which is so small as to be unobservable at any current or anticipated experiments would completely eliminate any constraints
from neutron star observations.

We see that these bounds can be tightly constraining for the case of bosonic dark matter, especially in the low-mass
region which would be relevant for the data of DAMA, CoGeNT, CRESST and CDMS.  But for this to be the case, the
stabilizing symmetry must also be a complexifying symmetry (thus forbidding self-annihilation); if it is broken to a parity, it must
be broken extremely weakly.  Likewise, these constraints are only relevant if self-interactions are either non-existent or
attractive.

If we had instead chosen
$\rho_X \sim 0.3~\gev /\cm^3$, the constraint on $\sigma_{nX}$ would be weakened by a factor $\sim 3000$.  Interestingly,
asymmetric dark matter can also be constrained by its effect on stellar evolution~\cite{StellarEvolution}.  In particular, a large density
of asymmetric dark matter in the core of the sun can lead to a change in the expected solar neutrino flux which differs from
observation.  Since the ambient dark matter density near the solar system is much less uncertain than in globular clusters, this
type of analysis of solar evolution provides a nice complement to bounds based on old neutron stars in globular clusters.

\subsection{Detection Strategies and Sensitivities}

For $m_X \lsim {\cal O}(10)\gev$, current direct detection experiments tend to rapidly lose sensitivity.  For such
models collider-based searches may provide the best possibility for detection.  Colliders can search for direct
dark matter pair production, through the same contact operator which mediates dark matter-nucleon scattering
interactions~\cite{MonojetTheory,Kumar:2011dr}.  These ``mono-anything" searches will produce missing transverse
momentum from the dark matter pair, along with
some radiated Standard Model particles which recoil against the dark matter pair.
Alternatively, colliders can produce heavy exotic particles which are charged both under $SU(3)_{QCD}$
and under the dark matter stabilizing symmetry.  The cascade decay of these heavy exotic particles will produce
Standard Model jets, plus missing transverse momentum.

An advantage of direct collider searches is that they probe the same
effective contact operators which mediate direct detection, thus potentially allowing one to correlate data from the
two detection strategies.  An advantage of cascade searches, however, is that they are based on the production of
QCD-coupled particles, a task at which hadron colliders excel.  But in the context of asymmetric dark matter, an
advantage of both search strategies is that they increase in sensitivity for low-mass dark matter, since it is easier
to produce low-mass dark matter at a collider.

For asymmetric dark matter at the lower end of the expected mass range, colliders
may thus provide the best sensitivity.  Future direct detection experiments may also begin to become sensitive to this region
of parameter space, but only for detectors using lighter target materials (to maximize the recoil energy arising from
the scattering of a low-mass dark matter particle) and for detectors with lower recoil energy thresholds.

Asymmetric dark matter in the $5-20~\gev$ range could potentially explain the data from DAMA, CoGeNT, CRESST and
CDMS.  These low-mass signals are in some tension with bounds from XENON100~\cite{XENON}, and there has much recent
work on potential resolutions of this tension arising from deviations from typical assumptions about dark matter
interactions and astrophysical distributions~\cite{LowMassReconciliation,IVDM}.  Most of these results also hold for
asymmetric dark matter models.

Asymmetric dark matter models which are relevant for the low-mass direct detection signals
are tightly constrained by LHC monojet searches~\cite{LHCmonojet}.  However, the sensitivity of such searches
depends in detail on the spin of the dark matter particle, the choice of effective operator, as well as the flavor
structure of the quark couplings.  Different
choices can lead to a dramatic weakening of these constraints, allowing consistency between the low-mass direct
detection signals and LHC bounds~\cite{Kumar:2011dr,Feng:2013vod}.
Moreover, if the particle mediating the interaction
has a relatively small mass ($\lsim 1~\gev$), then the scattering interaction may still be short-ranged, while the
effective operator approximation will not be valid for dark matter production at the LHC.  For such models, monojet
signals may be significantly suppressed, though it may be possible to directly produce the mediating particle at
colliders~\cite{Shoemaker:2011vi}.

Although asymmetric dark matter may potentially explain the low-mass direct detection signals, it cannot explain
the potential gamma-ray excess from the galactic center~\cite{GalacticCenter}.  Although this excess could be consistent with low-mass
dark matter, an annihilation cross section large enough to produce these signals would erase any dark matter
asymmetry.

\section{Outlook}

It is interesting to consider how one may distinguish asymmetric dark matter from a more standard WIMP candidate.
As we have seen, the main features one would expect from asymmetric dark matter are a low-mass particle with
no indirect detection signatures.  However, such a signature could also be reproduced by a particle with
$p$-wave suppressed annihilation.  In this case, a distinguishing signature may arise from dark matter searches
at neutrino detectors.

Neutrino detectors search for the neutrinos which arise when dark matter annihilates in the core of the sun.
If the sun is in equilibrium, then the rate at which dark matter is captured is the
same as the rate at which it is annihilated.  The key point here is that, if the sun is in equilibrium, then
the annihilation rate is independent of the annihilation cross section because the annihilation rate is equal to
the capture rate, which is determined by the scattering cross section.  In particular, a very small annihilation
cross section would imply a very large equilibrium density.  As a result, even dark matter with a very small
annihilation cross section can still
yield a detectable neutrino rate, provided the dark matter is in equilibrium.

If dark matter has a small mass, the ambient number
density will be very large.  As a result, dark matter with a scattering cross section large enough to
explain the low mass direct detection data could be in equilibrium even if $\langle \sigma_A v \rangle \lsim
0.01~\pb$~\cite{SunEquilibrium,Kumar:2012uh}.  As a result, low-mass dark matter with a largely $p$-wave annihilation cross section
and very small $s$-wave contribution (perhaps arising from chirality-suppressed annihilation to $b$-quarks or
$\tau$-leptons) can still yield a large event rate at neutrino detectors.

On the other hand, if low-mass dark matter is found using direct and collider search strategies, but
low-energy events are not seen at neutrino detectors, this implies that the annihilation
cross section is indeed very small.  While by no means determinative, this may perhaps provide evidence
suggesting that dark matter is asymmetric, with an annihilation cross section which is more heavily
suppressed than $p$-wave/chirality suppression.

%%%%%%%%%%%%%%%%%%%%%%%%%%%%%%%%%%%%%%%%%%%%%%%%
%% BACKMATTER
%%%%%%%%%%%%%%%%%%%%%%%%%%%%%%%%%%%%%%%%%%%%%%%%

\begin{theacknowledgments}
We are grateful to J.~Bramante, K.~Fukushima and E.~Stopnitzky for
collaboration.  We are also grateful to the organizers of PPC 2013.  Our attendance
at PPC 2013 was supported in part by Department of Energy grant DE-FG02-13ER41913.
\end{theacknowledgments}



\bibliographystyle{aipproc}   % if natbib is available

%\bibliography{sample}

%  \typeout{******************************************}
%  \typeout{** Please run "bibtex \jobname" to optain}
%  \typeout{** the bibliography and then re-run LaTeX}
%  \typeout{** twice to fix the references!}
%  \typeout{******************************************}

\end{document}